# Room-temperature ultrafast non-linear spectroscopy of a single molecule

Matz Liebel[1], Costanza Toninelli[2,3] and Niek F. van Hulst[1,4]

[1] ICFO-Institut de Ciencies Fotoniques, The Barcelona Institute of Science and Technology, 08860 Castelldefels, Barcelona, Spain.

[2] CNR-INO, Istituto Nazionale di Ottica, Sesto Fiorentino, Italy.

[3] LENS, Università di Firenze, Sesto Fiorentino, Italy.

[4] ICREA-Institució Catalana de Recerca i Estudis Avançats, 08010 Barcelona, Spain.

**Single molecule spectroscopy aims at unveiling often hidden but potentially very important contributions of single entities to a system's ensemble response. Albeit contributing tremendously to our ever growing understanding of molecular processes the fundamental question of temporal evolution, or change, has thus far been inaccessible, resulting in a static picture of a dynamic world. Here, we finally resolve this dilemma by performing the first ultrafast time-resolved transient spectroscopy on a single molecule. By tracing the femtosecond evolution of excited electronic state spectra of single molecules over hundreds of nanometres of bandwidth at room temperature we reveal their non-linear ultrafast response in an effective 3-pulse scheme with fluorescence detection. A first excitation pulse is followed by a phase-locked de-excitation pulse-pair, providing spectral encoding with 25 fs temporal resolution. This experimental realisation of true single molecule transient spectroscopy demonstrates that two-dimensional electronic spectroscopy of single molecules is experimentally in reach.**

Our fundamental understanding of key processes such as vision, light harvesting, singlet fission or the effect of coherences on molecular reactivity has dramatically benefited from the advent of non-linear ultrafast measurement techniques[1–4]. The convenient way to study these processes is transient absorption, or pump-probe, spectroscopy[5]. Here, a temporally short laser pulse photo-excites the system of interest and a time-delayed probe pulse reports on the spectro-temporal evolution on the excited electronic state of interest (Figure 1). Such experiments are routinely performed on molecular ensembles and sufficiently high signal levels are ensured by adjusting the molecular concentration within the probe volume. If one were to perform such an experiment on a single molecule, the sole optimization possibility is the reduction of the probe volume down to the diffraction limit. As a result, the illuminated area ($S$) at ambient temperature is, even in the ideal case, six orders of magnitude larger than a typical molecular absorption cross section ($\sigma$) or, in other words, only one in $10^6$ photons is absorbed by the single molecule thus resulting in a signal-to-background ratio of less than $10^{-6}$ (Figure 1). Given these considerations it is not surprising that even the comparatively easy task of detecting a single molecule in linear absorption has only been achieved recently and under great experimental efforts[6–8]. More challenging pump-probe experiments in resonance with a stimulated emission transition have so far, amid showing promising results, remained unsuccessful[9]. Under cryogenic conditions the absorption cross section increases dramatically but the dramatic reduction in homogeneous linewidth results in the loss of all dynamical information on femtosecond timescales[10,11]. In summary the only feasible possibility to resolve ultrafast transient dynamics of a single molecule is to capture the effect of the non-linear pump-probe process in a background-free fashion by fluorescence detection[12–14].

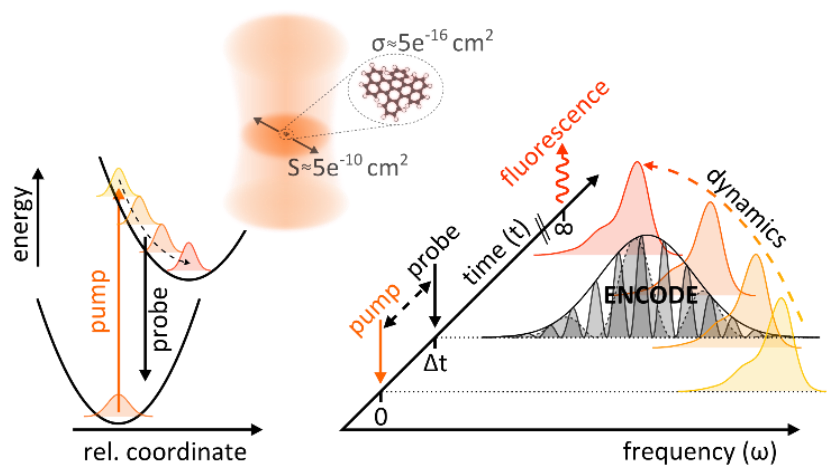

**Figure 1, Concept of transient absorption spectroscopy and single molecule implementation.** Schematic of a typical time resolved pump-probe experiment and signal considerations due to the diffraction limit for a possible single molecule implementation (left). Proposed experimental implementation of a broadband single molecule transient absorption experiment relying on fluorescence detection (right). The spectro-temporal dynamics of the excited electronic state are obtained by spectrally modulating (encoding) the probe pulse in order to modify the resulting fluorescence signal.

Here, we introduce transient ultrafast encoded single molecule spectroscopy (trueSMS) which is capable of resolving the transient spectral evolution of a single molecule on its excited electronic state with a temporal resolution of 25 fs at room temperature in condensed phase. TrueSMS relies on encoding[15–18] the spectro-temporal effect of a pump-probe experiment, with femtosecond precision, into a fluorescence signal, which is typically obtained nanoseconds later (Figure 1). In brief, a reflective Schwarzschild objective focuses the collinear pump and probe pulses onto a sample containing individual dibenzoterrylene (DBT) molecules embedded in a crystalline anthracene matrix (Figure 2a and Methods)[19]. A high numerical aperture (oil immersion) objective collects the fluorescence signal, which is, after passing a band-pass filter, imaged onto an EMCCD camera (Methods). By integrating the area of the fluorescence signal, we extract the absolute photon emission per integration time. In this simple implementation the experiment is capable of recording the transient change in fluorescence as a function of pump-probe delay, which is directly reporting on the dynamical evolution of the stimulated emission transition within the probe's spectral window, yet so far spectral resolution is missing.

To extract spectral information, we rely on Fourier transform spectroscopy. Here, the spectrum of interest is measured by interrogating the system with varying, but precisely defined, interferograms (Figure 2b). Commonly the interference between two pulses of a phase-locked pulse-pair yields the desired spectrum. Instead, we have chosen for pure amplitude-only pulse shaping to mimic these effective time delay dependent interferograms. As a more generalized concept, the interferogram method can be interpreted as a way of experimentally enforcing a set of sinusoidal basis functions. Under these conditions, a Fourier transformation is able to extract the desired spectrum from the measured data. Any other spectral modulation approach is, however, equally well suited albeit requiring a different decomposition or generation method (Supplementary Information 2). For example, compressed sensing might employ random sampling whereas commercial absorption spectrometers scan a spectrally narrow line across the frequency range of interest [20]. Irrespective of the method, if a pulse is resonant with a ground to excited electronic state transition of a fluorescent molecule it is possible to obtain its fluorescence excitation spectrum by employing any of the mentioned techniques with the interferogram methods allowing for a direct decomposition by Fourier transformation[21–26].

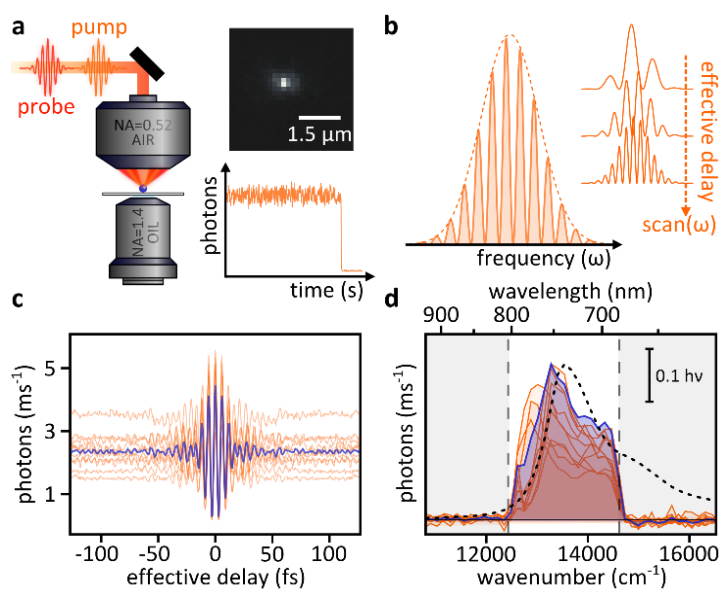

**Figure 2, Experimental implementation of spectral modulation spectroscopy and proof of concept experiments on single molecules.** (a) Experimental implementation of a single molecule transient absorption experiment and representative images of a single molecule and its single-step bleaching event. (b) Concept of spectral measurements by spectral amplitude modulation of the probe pulse, mimicking the spectral control obtained from a phase-locked pulse-pair. (c) Fluorescence traces recorded by pure spectral modulation of the excitation pulse for individual single DBT molecules (orange) as well as the averaged signal (blue) using a carrier frequency of 19231 cm$^{-1}$ [20]. (d) Fluorescence excitation spectra obtained from (c) as the real part of the fast Fourier transformation of the respective fluorescence traces for the individual molecules (orange) and the average of all traces (blue). An ensemble absorption spectrum of DBT dissolved in Toluene (dashed line) is shown for comparison. The white area indicates the spectral window of the excitation pulse, the regions outside of this detection window are shown to allow for an estimate of the absolute noise contributing to the measured spectra. (Supplementary Information 1).

To measure the spectrum of an excited electronic state it is necessary to first photo-excite the system with a pump pulse and then modulate the spectrum of the probe pulse resonant with a stimulated emission transition as described previously. The effect of the spectral modulation of the probe pulse on the fluorescence signal yields, after Fourier transformation, a fluorescence de-excitation, or stimulated emission, spectrum at the specific pump-probe delay. Electronic dephasing times at room temperature are approximately 50 fs, hence it is necessary to scan the effective time delay over a similar temporal range to obtain a sufficiently resolved spectrum after Fourier transformation [27]. Our choice for interferogram generation by pure amplitude shaping yields a temporal envelope reminiscent of a physical pulse-pair, albeit small satellites of approximately 10% of the central pulses intensity being generated (Supplementary Information 2). Given the limited signal-to-noise ratios achievable in trueSMS experiments the satellite contribution to the total signal is, however, outside of our detection limit. The spectral modulation necessary for amplitude shaping of an effective pulse-pair delay is computed as described previously[26,28]. Additionally, we adjust the carrier frequency, to ease the required spectral resolution needed for pulse shaping, which is analogous to performing the experiment in a rotating frame, a concept routinely used in nuclear magnetic resonance (NMR) spectroscopy (see Methods) [26,29]. Importantly, even though our approach might at first seem to violate the uncertainty principle, the light-matter interaction induced molecular response and hence the recorded signal remain within Fourier limited boundaries as discussed in detail in the context of impulsive vibrational spectroscopy [30,31]. In other words, the ultimate time-resolution of our experiment, or any other experimental approach, is either determined by the pump-probe cross correlation or, if spectral information is desired, by the convolution of the former with the electronic decoherence time, or spectral bandwidth, of the molecular transition under observation as discussed in detail by Mukamel *et al.* [30] as well as in the Supplementary Methods 2,3.

To demonstrate the validity of this approach we record the fluorescence signal of multiple but individual DBT molecules while spectrally modulating the excitation pulse according to effective delays ranging from 0 to 126 fs. We mirror the trace to obtain the full interferograms (Figure 2c) which directly yield the individual fluorescence excitation spectra as the real part of their respective Fourier transformation (Figure 2d and Supplementary Information 4). The ensemble absorption spectrum of DBT dissolved in toluene shows good qualitative agreement with the averaged fluorescence excitation spectrum, whereas the variation between individual spectra highlights the dependence of vibronic transitions on the dielectric environment of the molecule[32]. Importantly, even though these spectra are obtained with a constant spectral phase, the level of agreement with the ensemble measurement is comparable to results obtained with pulse-pairs, thus validating our experimental approach[25,26].

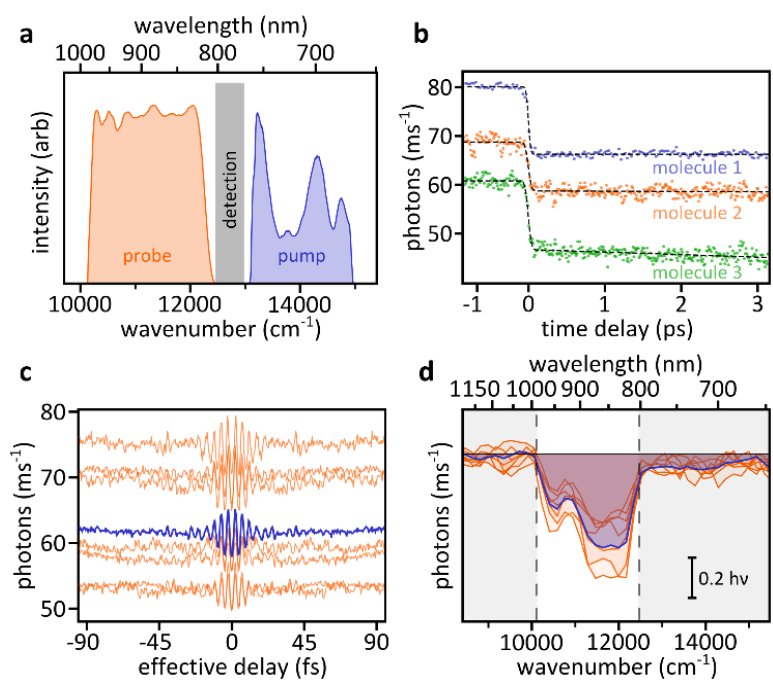

**Figure 3, Single molecule transient fluorescence spectroscopy and transient stimulated emission spectra.** (a) Spectra of the pump (blue, 17.6 fs) and probe (orange, 15.0 fs) pulses employed in the trueSMS experiments as well as the fluorescence detection window (black). (b) Representative transient fluorescence traces recorded for three different single DBT molecules (dots) alongside their fits to a kinetic model composed of a step function convolved with a Gaussian rise time and a single exponential decay (dashed line). (c) Fluorescence traces recorded by trueSMS for individual single DBT molecules (orange) as well as the average of all traces (blue). The traces are recorded at a pump-probe time-delay of 5 ps and a carrier frequency of 19231 cm$^{-1}$. (d) TrueSMS spectra obtained from (c) as the real part of the fast Fourier transformation of the respective fluorescence traces of the individual molecules (orange) and the average of all traces (blue). The white area indicates the spectral window of the probe pulse employed in the experiment. The apparent shoulder in the averaged spectrum outside of the detection window (<800 nm) is discussed in detail in Supplementary Information 5.

After these proof-of-concept demonstrations, we are in a position to perform trueSMS experiments on individual DBT molecules. We employ a pump pulse that almost fully covers the ground state absorption spectrum of DBT and a probe, or depletion, pulse that overlaps with most of the stimulated emission transition while only leaving a narrow window for fluorescence detection (Figure 3a and Methods). A scan of the pump-probe time-delay reveals a partial depletion of the fluorescence signal of typically 20%, which varies for different molecules. The temporal response of the de-excitation step is about 40 fs, which is slightly larger than the measured cross correlation between the pulses of 28 fs, and is most likely due to the non-instantaneous Stokes shift of the system (Figure 3b). Under our experimental conditions, the level of fluorescence depletion scales linearly with the probe power, which is crucial for the success of the following trueSMS experiments (Supplementary Information 5).

To obtain trueSMS spectra we choose a pump-probe delay of 5 ps and modulate the probe pulse to achieve spectra corresponding to effective delays ranging from 0 to 96 fs, while recording the fluorescence signal. Figure 3c shows the respective interferograms obtained for several individual DBT molecules. An important difference compared to the ground state excitation experiment (Figure 2c) is that the depleted fluorescence signal exhibits a minimum at time zero, which indicates that the modulated pulse reduces the overall signal as expected for a stimulated emission transition.

Figure 3d shows the corresponding de-excitation, or stimulated emission spectra, which are directly obtained as the real part of the Fourier transformation of the respective interferograms. A result of the π phase shifted interferograms is the negative spectral amplitude, thus directly indicating that the action of the probe pulse results in a loss of fluorescence intensity. The non-zero signal observed in the average spectrum between 650-800 nm is most likely a minor artefact caused by pulse shaping in the near-infrared (Supplementary Information 6).

Finally, we resolve the ultrafast spectral dynamics of a single DBT molecule. To this end, we employ the same pulses as previously, but focus on the first tens of femtoseconds after photoexcitation (Figure 4a). We select six pump-probe time-delays within ±100 fs around time-zero, to capture the spectral evolution of the stimulated emission spectrum during and immediately after interaction with the pump pulse. One additional time-delay at 3.5 ps is chosen to obtain a reference spectrum close to that of a vibrationally cold molecule. Figure 4b shows signals recorded as previously but now at the seven different pump-probe delays alongside the signal obtained under identical conditions after single-step photobleaching of the molecule. We remark that all seven traces as well as the background were measured on the same single molecule, or at the exact same sample positon in the case of the background.

The real part of a Fourier transformation reveals the ultrafast spectral evolution of the single molecule with only minor contributions before time-zero (Figure 4c). Importantly, neither the Anthracene matrix nor any other part of the sample contribute to the reported signal as underlined by the flat spectral background obtained after photobleaching. We observe an initially uniform rise of the stimulated emission signal followed by a pronounced amplitude gain at the blue side of the spectrum. Only in the long time-delay spectrum do we observe the two-band feature reminiscent of a vibrational progression that is also present in the spectra reported previously at a time-delay of 5 ps (Figure 3d). The observation of such a progression might, at first, seem surprising as a fluorescence spectrum reported previously appears almost featureless[19]. Those data, however, report predominantly on the blue spectral region of the emission spectrum whereas our experiment explores the stimulated emission transition much further into the near-infrared. Both the absence at early pump-probe delays as well as the appearance at longer delays reminiscent of vibrational cooling was repeatedly observed during several weeks of experiments under varying experimental conditions. Especially the spectral evolution makes us believe that the feature reported here is due to a vibronic transition of the single molecule.

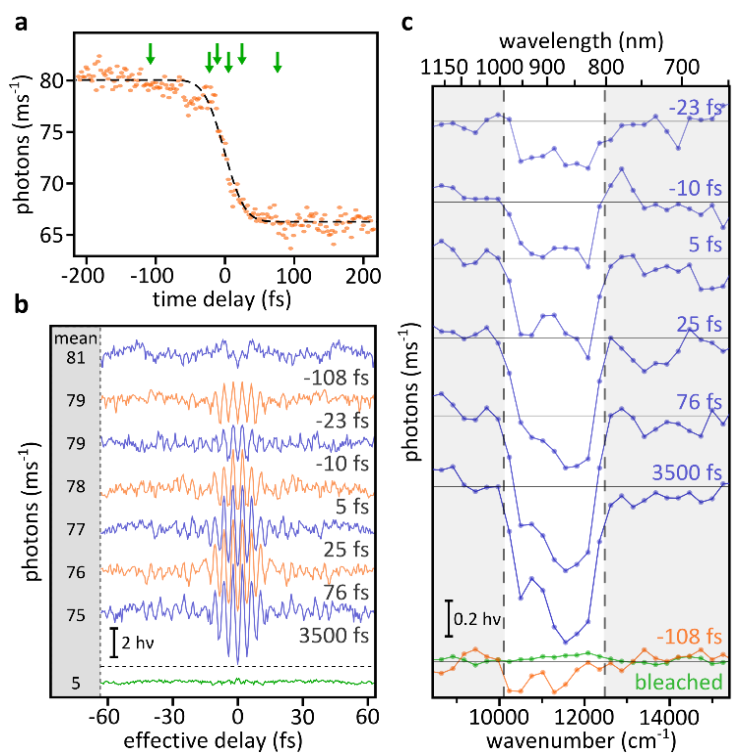

**Figure 4, Ultrafast broadband single molecule transient absorption spectroscopy.** (a) Transient fluorescence trace of a single DBT molecule (dots) and its fit to a kinetic model composed of a step function convolved with a Gaussian rise time and a single exponential decay (dashed line). The green arrows indicated the points of trueSMS measurements. (b) Fluorescence traces recorded by trueSMS for the same DBT molecule as in (a) at the indicated pump-probe delays (orange and blue). The traces are offset for clarity; the respective mean fluorescence levels are indicated on the left. The green trace shows the background signal recorded after one-step photobleaching of the DBT molecule took place highlighting the absence of effective time-delay induced oscillations as well as the drastically reduced mean signal (5 vs. 80 photons per ms). (c) TrueSMS spectra obtained as the real part of the fast Fourier transformation of the respective traces (blue) after subtraction of the residual pre time-zero background. All spectra are offset for clarity with the solid black lines indicating zero amplitude of the respective spectrum. The pre time-zero spectrum (orange) as well as the spectrum obtained after photobleaching (green) are included for clarity.

In summary, we report the first transient stimulated emission spectrum of a single molecule as well as the first direct observation of the dynamical evolution of such a spectrum on an ultrafast (<30 fs) timescale in condensed phase at room temperature. We are able to resolve spectro-temporal dynamics, which we attribute to a combination of Stokes shift as well as vibrational relaxation. Our methodology relies on our ability to encode precise spectral information at a well-defined pump-probe time-delay and to retrieve this information by means of fluorescence at a much later point in time. A combination of amplitude only pulse shaping with concepts borrowed from NMR spectroscopy allows us to convert the encoded information into an excited state emission spectrum at the respective time-delay while, simultaneously, keeping the temporal resolution of the experiment close to the Fourier limit as the pump-probe cross correlation only marginally broadens this electronic decoherence time dominated limit. The spectral encoding approach allows us to access molecular dynamics of single molecules within the first few tens of femtoseconds after photoexcitation. Even conventional transient absorption experiments, performed at the ensemble level, struggle to access the 10s of femtoseconds time scale due to nonlinear signals being generated by the temporally overlapping pump and probe pulses[33]. At the single molecule level, advantageously, it is possible to circumvent this limitation. The signal arising from the single molecule itself shows a distinctly different spatial emission pattern as compared to all nonlinear signals generated in the sample matrix (or support) and a simple dark-field mask in the back-focal-plane of the collection objective efficiently eliminates their contributions[34,35]. This notion is further supported by the complete absence of spectral amplitude for the experiment performed after photobleaching (Figure 4c). Importantly, this advantage is not due to a somehow superior temporal-resolution, which would require a violation of the Fourier limit. The distinct advantage merely stems from the complete elimination of all parasitic non-linear signals, which often render ultrafast dynamics on the fastest timescales invisible irrespective of the temporal resolution. This quasi background free detection allows trueSMS to precisely determine the spectral means of the evolving molecular system at any given pump-probe delay, with femtosecond precision, reminiscent of super-resolution microscopy whereas ensemble based methods are only benefiting from this possibility at later time scales (see Supplementary Information 3 for signal simulations).

The trueSMS experiment presented here is conceptually almost identical to fluorescence detected 2D electronic spectroscopy albeit the latter two field contributions not being physically separated[17,36–38]. As such, only probe interactions leading to a modulation of the total fluorescence intensity can be detected. In its current form, trueSMS is sensitive to both ground state bleach and stimulated emission signals as both result in a reduction of the signal. Molecular transitions such as an excited state absorption followed by rapid internal conversion to the emissive excited electronic state in the spirit of "*Khasha's Rule*" are invisible. Even though this fact might, at first, appear to be a disadvantage we believe that the resulting simplification of the resulting dynamics might allow less error prone interpretation of the underlying molecular dynamics. Experiments, such as pump-push, that allow for state or pathway specific observations have tremendously contributed to our understanding of molecular processes, especially in the context of semiconductor research[1,39,40]. TrueSMS operates in the spirit of such approaches and we envision combined fluorescence/photocurrent experiments that allow for the firm assignment of loss vs. charge-separation channels in the future.

In summary, trueSMS addresses both ground and excited electronic states by spectral shaping, and even 2D information can be retrieved when combined with advanced signal acquisition as for example the concept of compressed sensing[20]. We therefore believe that our initial steps will ultimately enable 2D trueSMS spectroscopy for single emitters such as quantum dots, plasmonic structures and ultimately single molecules[41].

## Methods

**Sample preparation,** The DBT@Antracene samples are prepared according to a protocol published previously [19] by spin-coating a 10 pM DBT in Toluene/Anthracene solution to obtain DBT doped anthracene crystals on No. 1.5 cover-glass. Fresh samples are prepared on a daily base to eliminated problems associated with Anthracene sublimation.

**Optical setup and measurements,** The output of an Octavius-85M (~6 fs pulses, 650-100 nm, 85 MHz) is pre-compressed by means of second- and third-order compensating chirped mirrors (Layertec GmbH) before being send into an all-reflective home build 4f grating stretcher operated in transmission geometry equipped with NIR polarisers. A spatial light modulator (SLM; Jenoptik SLM-S640d) is placed into the Fourier plane alongside hard aperture filters to select the spectral envelope of the pump and the probe pulses. Pump and probe are separated by means of a dichroic laser beam-splitter (HC BS R785/ AHF Analysentechnik AG). The pump passes a mechanical delay line before being recombined with the probe pulse using the same beam-splitter.

The collinear pulses are send into a home build transmission microscope using a 0.52 numerical aperture reflective objective (36x, 50102-02 Newport) for illumination and a 1.4 numerical oil-immersion objective (CFI Plan Apo Lambda 60X Oil, Nikon) for collection. A 3 mm diameter aperture stop is mounted in the vicinity of the back-focal-plane of the collection objective to remove phase-matched third-order signals generated in the sample support as well as photons present in the detection window that were not perfectly removed by the hard aperture spectral filtering. After passing a 570 nm long-pass filter (570ALP, Omega Optical) and two 22 nm band-pass filters centred at 786 nm (786/22 Brightline, Semrock) the signal is imaged onto an emCCD camera (ImagEM X2 EM-CCD camera, Hamamatsu) by means of a 300 mm focal length achromatic lens. The imaging system exhibits a magnification of 97 corresponding to 165 nm/px. The illumination spot size has a diameter of 1.0 µm at FWHM. Both pump and probe exhibit the same, linear, polarisation and a half-wave-plate is used to align the polarisation with the transition dipole moment of the individual molecules. All experiments are performed in the linear regime with respect to both pump and probe pulse (Supplementary Information 5 for details).

**Pulse compression,** Pulses are compressed with the SLM and a 10 µm thick β-barium borate crystal using multiphoton intrapulse interference and second harmonic frequency resolved optical gating [42,43].

**Spectral measurements and corrections,** All spectra are measured in the focus of the Schwarzschild objective by collecting scattering from non-resonant sub diffraction limited particles as mentioned above but without the filters while scanning either a pulse-pair delay or a spectrally narrow band-pass filter on the SLM. All spectra are corrected for the wavelength dependent scattering amplitude as well as the quantum efficiency of the camera and all optics in the collection path. These measurements are then used to compute the appropriate amplitude masks in order to generate the square-amplitude spectra; the successful spectral shaping is verified with a second measurement (Supplementary Information 1 for details).

**Pulse shaping,** To compute the effective delay dependent pulse intensity we employ the following equation:

$$I(\nu, \Delta t) = |\cos[\pi(\nu - \nu_0)\Delta t]|^2,$$

with $\Delta t$ being the effective time delay, $\nu$ the optical frequency and $\nu_0$ the carrier frequency. The computed modulation is then applied to the SLM, which exhibits an approximate spectral resolution of 10 cm$^{-1}$/px. No phase modulation is applied beyond the one necessary for temporal fine compression of the pulses as obtained previously by multiphoton intrapulse interference.

**TrueSMS measurements,** TrueSMS experiments are performed by averaging 4 images (50 ms per image) followed by changing the SLM mask to the next effective delay thus requiring 300 ms per time point. A total of 121 time points are acquired for each effective delay trace thus requiring a total acquisition time of 36 s per pump-probe delay or 4 min per full experiment. Scans are repeated and averaged until photobleaching while rejecting large intensity jumps caused by photobleaching, which requires the acquisition of multiple full trueSMS scans. The data reported here has been obtained by averaging 15 spectral runs but we remark that similar results were obtained with 2-4 runs and 75 time points depending on the blinking behaviour of the molecule. Based on our experiments we estimate that fast detection with an avalanche photodiode in combination with a rapid scanning interferometer should enable qualitatively identical results within less than 5 min as blinking events can statistically be rejected from the data averaging.

# Supplementary Information

**Supplementary Information 1. Excitation pulse used for fluorescence excitation spectroscopy**

Prior to any spectral measurement by amplitude modulation it is necessary to determine the spectrum of the pulse employed. This concept holds even for commercial spectrometers where a spectrally narrow line is scanned across the spectrum and the transmission through a cuvette is measured. Once the measurement is complete a normalisation for the spectrally dependent line intensity yields the transmission (absorption) spectrum. Such a renormalisation can be relatively problematic in Fourier transform spectroscopy as a large portion of the frequency range obtained after Fourier transformation was "measured" without, or very little, intensity being present in the measurement pulse. A renormalisation hence yields very large or even infinite values. These non-measured spectral windows do, however, contain important information as they are directly reporting on the measurement noise hence allowing the reader to judge if spectral features within the measurement range (the spectral region of sufficient intensity of the measurement pulse) are statistically relevant i.e. above the noise level.

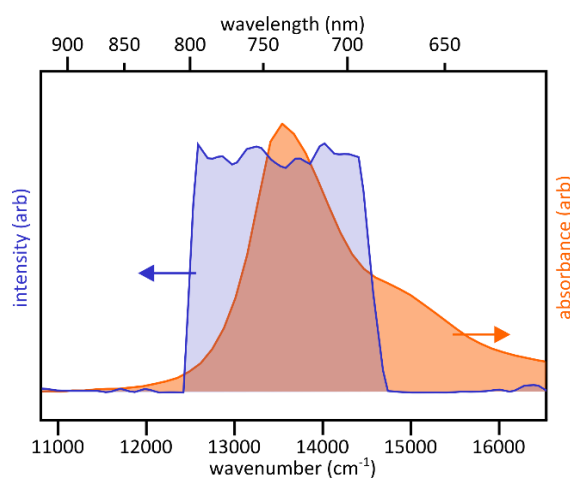

**Supplementary Figure 1:** Pump spectrum measured at the sample position, as scattering from a non-resonant sub-diffraction limited particle (blue) in comparison to the absorption spectrum of DBT dissolved in Toluene (orange).

To circumvent this limitation we decided to generate measurement pulses with square spectral envelopes by means of pulse shaping with the SLM. To adjust the spectrum we operate the microscope in darkfield configuration and image a sub diffraction limited scattering particle in the focus of the Schwarzschild objective. We then measure the pulse spectrum by scanning a spectrally narrow amplitude mask on the SLM thus generating a spectrally narrow band which is identical to the operation principle of most absorption spectrometers. From this measurement we are able to compute an amplitude correction that is consecutively applied to the SLM. The results obtained are verified by scanning an effective delay thus simulating the excitation spectroscopy, or trueSMS, experiment (Supplementary Figure 1 shows the pump pulse used for obtaining the data presented in Figure 2 of the main manuscript). After this adjustment we insert the necessary fluorescence detection filter and begin the measurements on DBT under otherwise identical conditions. Even though the spectra obtained are not perfectly square pulses their residual spectral modulations are comparable to the measurement noise of the experiments presented in the main manuscript and hence do not negatively impact on our conclusions. One drawback of this method is that the trueSMS, or fluorescence excitation, spectra seemingly decay towards the edges of the measurement pulse. The alternative of renormalising does, however, not solve this issue as it leads to the opposite problem of dramatic amplification of the noise as well as the spectral amplitude[25]. In summary, we believe that the pre-normalisation of the measurement pulses is a highly beneficial approach to Fourier transform spectroscopy as it automatically yields correct spectral amplitudes while, simultaneously, allowing for a direct comparison with the experimental noise level.

**Supplementary Information 2. Temporal resolution differences between Fourier spectroscopy modalities**

**a) Pulse shaping and parasitic satellites,**

In the trueSMS experiments presented we opted for a rarely employed amplitude-only pulse shaping approach in order to encode the action of a physical pulse-pair into the probe pulse. To justify this approach we opt for a comparison of different pulse shaping approaches by means of spectral, temporal as well as Wigner representations of the respective electric fields encountered.

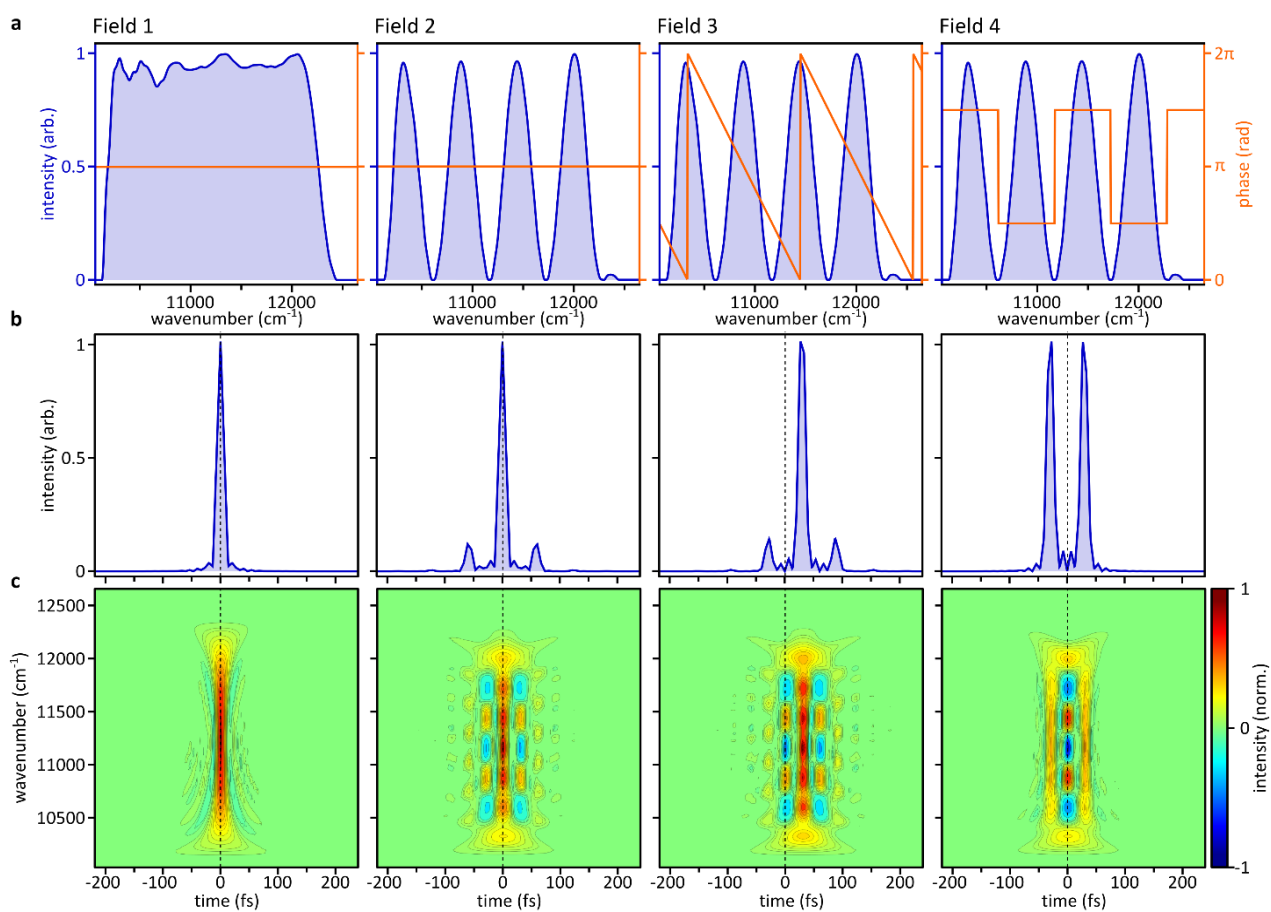

**Supplementary Figure 2: Pulse shaping and spectro-temporal electric field representations.** (a) (from left to right): original pulse followed by three pulses with an intensity modulation corresponding to a 60 fs pulse-pair (carrier frequency of 19230 $cm^{-1}$) and different phase terms. The flat phase has been applied in the experiments presented in the manuscript. (b) Corresponding temporal intensities. (c) Time-frequency, or Wigner, representations of the electric fields generated by the respective phase and amplitude modulations. The positive (red) and negative (blue) amplitudes can be interpreted as electric field components which are out of phase.

Supplementary Figure 2a shows the spectrum and phase of the unshaped pulse employed in all trueSMS experiments (Field 1) in comparison to a spectrum obtained by applying an amplitude mask corresponding to a pulse-pair of 60 fs separation and a carrier frequency of 19230 $cm^{-1}$. Three different phase modulations are applied: a flat phase (as in the trueSMS experiment; Field 2), a linear phase ramp (Field 3) as proposed by Shim and Zanni[37] and a binary 0-π phase step mask (Field 4) as often employed for generating pulse-pairs [26,28,44]. To gain further insight into the temporal interaction of the probe pulse-pairs with a molecule we consult the respective temporal intensity envelopes of these four pulses (Supplementary Figure 2b). Field 2 and 3 exhibit the same satellite pulses at +- 60 fs (inter-satellite distance 120 fs or 2xΔt) with the main intensity being present in the centre between the satellites. The sole difference between these approaches is that the latter is temporally shifted by 30 fs, or Δt/2, as a result of the linear phase term adding a 30 fs time delay. The temporal envelope of Field 4 is composed of two temporal envelopes shifted by +- 30 fs with respect to the original time-zero. The small satellite pulses observable at, the unintended distance of, twice the pulse-pair separation seen in field 2 (Supplementary Figure 2b) are generally disadvantageous for spectral measurements and field 4 therefore represents, in our opinion, the ideal pulse-pair. At our signal-to-noise ratios, however, the satellites' relative intensity of 10% with respect to the main band renders their signal contributions essentially invisible. Based on purely experimental considerations we opted for field 2 as we observe increased spectral leakage, in a pulse-pair separation dependent fashion, when employing shaping approaches exhibiting phase jumps such as field 4. As our experiment is operated without spectral clean-up filters this leakage results in parasitic background signals which reduce the overall data quality of the trueSMS experiments. Field 2 does not generate such pulse-pair separation dependent leakage and is therefore well suited under our experimental conditions.

**b) Wigner representation and light-matter interactions,**

The strict time or frequency representations discussed above are, not sufficient to understand the full light-matter interaction of the complex electric fields with a molecule of interest and we therefore introduce a Wigner representation which maps the electric field as function of both time and frequency in a pseudo "3D" fashion (phase is to some extend encoded in positive/negative amplitude).

Supplementary Figure 2c shows the Wigner representations of all four electric fields. Field 1 shows the expected time-frequency behaviour of a transform limited pulse. In comparison, fields 2 and 3 exhibit pronounced modulated side lobes, as expected for a pulse-pair with 60 fs delay. The alternating positive/negative amplitudes can be understood as electric field components with comparable amplitude but opposite, π-shifted, phase. The frequency of this modulation increases with increasing pulse-pair separation as intuitively expected from the increasing modulation frequency of a pulse pair spectrum. Field 4 shows qualitatively the opposite behaviour of the previous fields, with a modulated centre band and two non-modulated side lobes, which is not surprising given that the 0-π binary phase applied corresponds to the phase modulation observed at the side lobes of fields 2 and 3. When integrating the fields across their temporal dimensions the components will interfere in either a constructive (red vs. red) or destructive (red vs. blue) fashion thus yielding the spectra shown in Supplementary Figure 2a. The same is true for integration in the other dimension which yields the temporal envelopes shown in Supplementary Figure 2b.

If a molecule interacts with any of the electrical fields shown in Supplementary Figure 2 it essentially performs an integration over the spectral bandwidth determined by its molecular spectrum. In the case of the modulated, phase shifted, electric field components such an integration leads to constructive or destructive interference, depending on the modulation frequency or pulse-pair delay, and therefore to a modulation of the de-excitation probability which manifests itself in the oscillations observed in all Fourier transform spectroscopies. In the limit of very fast modulation frequencies (long pulse pair separation) the molecule is essentially blind to these parts of the electric field as destructive interference within the spectral bandwidth of the molecule results in zero de-excitation probability.

The non-modulated part, as for example the central band of field 2, still results in de-excitation. Importantly, the former interaction results in temporally varying, or oscillating, signals as the pulse-pair distance is scanned and the latter in a temporally constant signal. Fourier transform spectroscopy, and therefore trueSMS, isolates the non-stationary signal and is therefore insensitive to the constant contribution.

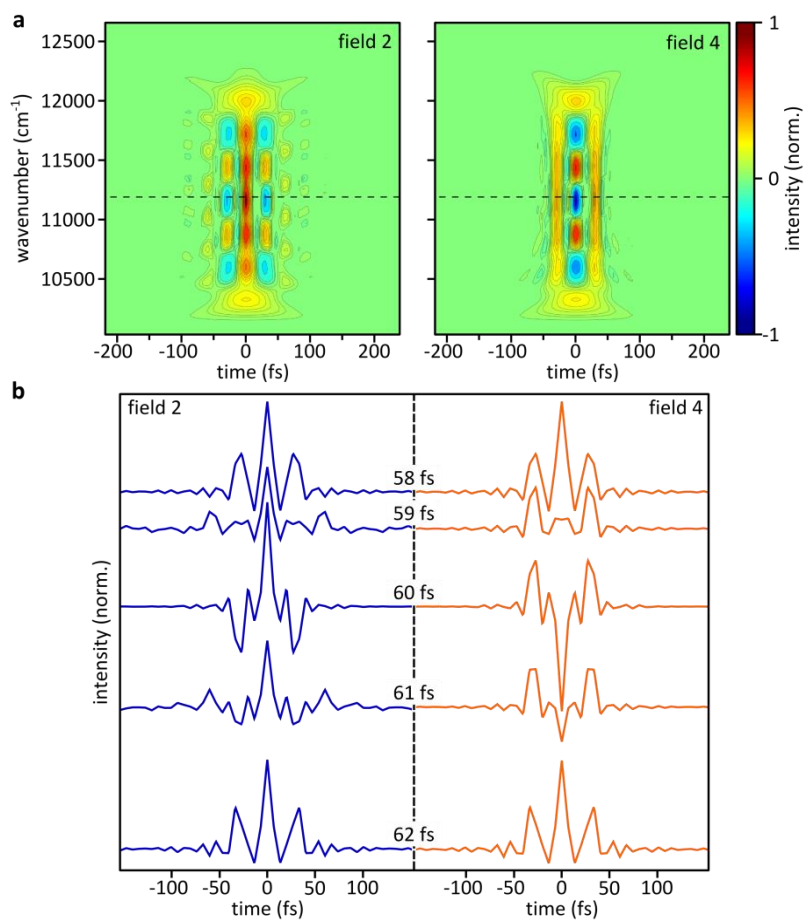

**Supplementary Figure 3: Pulse-pair separation dependence of electric fields.** (a) Wigner representations of field 2 and 4 at a pulse-pair separation of 60 fs as shown previously. The dashed lines at 11188 cm$^{-1}$ indicate the position of the temporal cut shown below. (b) Temporal evolution of the electric field at 11188 cm$^{-1}$, as a function of pulse-pair separation from 58 to 62 fs.

TrueSMS requires two electric field interactions with the probe pulse in order to convert excited state into ground state population. These interactions have to occur within a time-frame dictated by the electronic decoherence time of the molecule which is on the order of 50-60 fs at room temperature. The two molecular interactions with the electric field are "smeared" across this temporal window which is a direct manifestation of the Fourier limit. As a result field 2 and field 4 qualitatively exhibit the same temporal resolution even though it might appear as if the modulation being present at time zero results in an improvement. Broadly speaking, one molecular interaction with field 2 at -60 fs followed by a second at 0 fs yields the same de-excitation probability as interactions with field 4 and the contribution to the temporally modulated signal extracted by Fourier transformation is therefore identical for both fields. This notion becomes clearer when consulting Supplementary Figure 3, which compares representative spectral cuts of the Wigner representations for different pulse-pair separations of both field 2 and 4, showing that the satellites carrying the actual interferometric signal represent about 50% of the intensity of the central band, yielding the same contrast for both field cases.

For longer separations, the modulated signal vanishes as discussed previously. The difference between fields 2/4 and field 3 is that the former two maintain their spectral mean at the same temporal position whereas field 3 exhibits a temporally shifting mean. We define pump-probe delays as the distance between the temporal intensity means of the respective pulses and therefore opted for field 2.

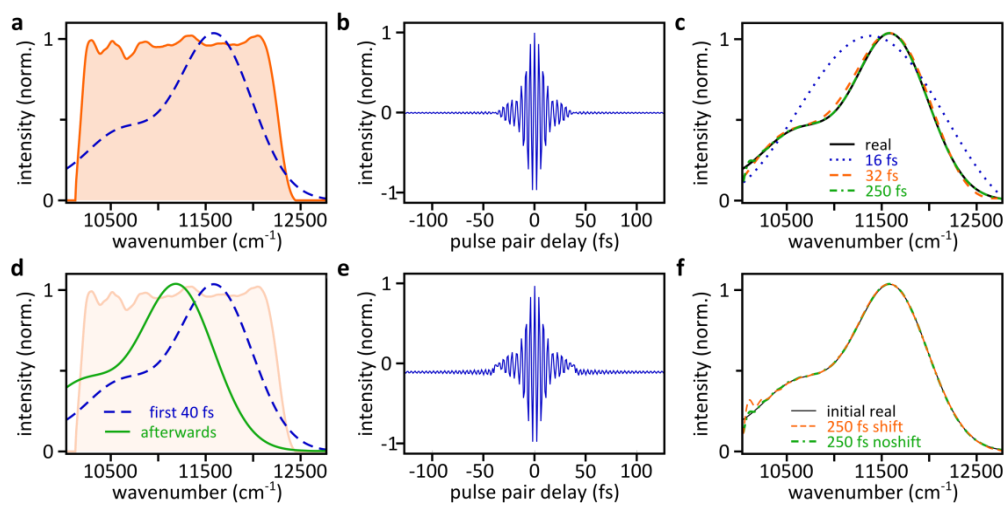

**Supplementary Figure 4: Time resolution and integration window.** (a) Excitation pulse spectrum (solid, orange) and molecular spectrum obtained by mirroring and shifting the DBT ground state absorption (dashed, blue). (b) Interferogram generated by computing excitation pulse spectra corresponding to a pulse pair with a carrier frequency of 19230 cm$^{-1}$, the trace has been truncated at ±125 fs for clarity. (c) Spectra obtained from truncated parts of (b) by means of Fourier transformation in comparison to the real spectrum. (d) As (a) but for a molecular system that abruptly changes its spectrum at a pulse-pair delay of 40 fs. (e) Analogous to (b) but for the evolving spectrum. (f) Spectra obtained by Fourier transformation of (b) and (e) in comparison to the initial spectrum.

To illustrate this rather abstract treatment in a more intuitive way we perform some simple Fourier transform spectroscopy simulations which are presented in Supplementary Figure 4. To generate experimentally relevant parameters we employ the same pulse spectrum as in the experimental implementation of trueSMS and a molecular spectrum that we generate by mirroring and spectral shifting of the ground state absorption spectrum of DBT (Supplementary Figure 4a). We generate the expected interferogram by computing the overlap integral of the respective spectra using 1 fs pulse-pair steps and a carrier frequency of 19230 cm$^{-1}$ over a total of ±250 fs (Supplementary Figure 4b). Qualitative examination of the interferogram shows that most of the temporally modulated information is present close to time-zero. To quantitatively compare the spectral information content we truncate the data at different pulse-pair delays, zero-pad the resulting pulse-pair traces to yield vectors of identical length, and compare the resulting spectra obtained after Fourier transformation with the original spectrum (Supplementary Figure 4c). Not surprisingly, already a relatively short pulse-pair delay of ±32 fs reproduces the spectrum almost perfectly whereas the ±16 fs cut merges the two spectral bands as a result of the Fourier limit. Increasing the integration window to ±250 fs only results in, at the signal-to-noise levels of our trueSMS scans, cosmetic improvement over the ±32 fs spectrum. Importantly, these observations are not a result of zero-padding which has no effect on the resolvability or information content of spectral data [45].

We conclude with a simulation demonstrating that excessive pulse-pair delays do not deteriorate temporal resolution which is only limited by the spectral response, or electronic decoherence time, of the system under study as well as the pulse durations employed. We simulate a scenario of a constant molecular spectrum that abruptly changes its position at a pulse-pair delay of 40 fs and its resulting interferogram as discussed previously (Supplementary Figure 4d,e). As a result of the changing overlap integral a pronounced step-function is observed at ±40 fs. Fourier transformation of this interferogram produces a spectrum reminiscent of the original, non-shifted, spectrum with only slight distortions at the red wing of the spectrum (Supplementary Figure 4f). This observation is not surprising given that the spectra obtained over a ±32 fs and ±250 fs pulse-pair window are essentially identical. Coming back to the previous discussion of the Wigner representations, the information content is predominantly contained within the temporal window dictated by the spectral width, or electronic decoherence time, of the molecular system and beyond this window the destructive interference between out-of-phase components of the electric field results in a close-to-zero contribution to the frequency resolved spectrum.

Based on these simulations we estimate our absolute temporal resolution of the trueSMS experiments to being essentially decoherence time limited. If we assume stokes shifted absorption and emission spectra which is, in most cases, a justified assumption we are able to use the simulation results extracted above in combination with the 28 fs instrument function to approximate our temporal resolution as $\sqrt{28^2 + 32^2} \approx 40$ fs even though we scan our pulse-pairs for longer delays. A reduction of the temporal duration of pump and probe pulses will therefore only marginally improve the achievable resolution as the molecular response is the dominating factor. Importantly, the pulse employed in the Fourier transform measurement has to be temporally compressed in order to measure the spectrum at the correct pump-probe delay for all frequencies. If this is not the case different parts of the spectrum will report on different time delays.

**Supplementary Information 3. Measuring temporal evolution on time-scales faster than the instrument response function**

To rationalise the spectral shifts observed in the trueSMS experiments which, apparently, occur on sup-temporal resolution time scales we perform simulations based on the estimated absolute temporal resolution of 40 fs (Supplementary Information 2). We simulate measurements on a spectrally evolving molecular system at time increments below this resolution analogous to the trueSMS experiment (Supplementary Figure 5). The system is composed of two Gaussian bands, reminiscent of a vibronic transition, with spectral amplitudes and centre frequencies as defined in Supplementary Figure 5a.

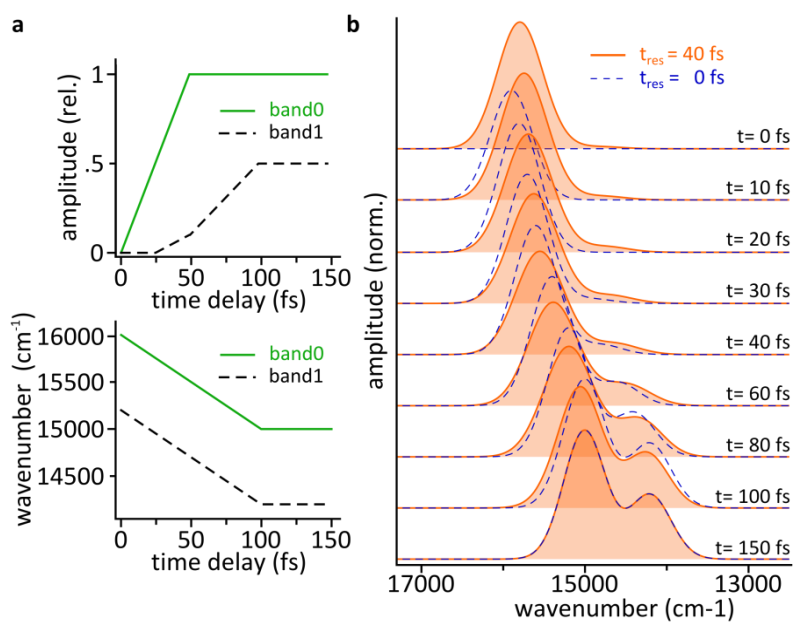

**Supplementary Figure 5: Measuring fast spectral evolution with limited temporal resolution.** (a) Time dependent amplitude (top) and central frequency (bottom) of the two vibronic bands of FWHM = 600 cm$^{-1}$ used to generate the stimulated emission spectrum. (b) Simulated transient stimulated emission spectra as measured with a time resolution of 40 fs at FWHM (orange, solid) or 0 fs (blue, dashed). The pump-probe delays are indicated on the right, all individual time-delays are normalised to the respective maximum amplitude for clarity.

Supplementary Figure 4b shows the expected spectral measurements with our finite resolution in comparison to the instantaneous spectrum at the specified time delay. To obtain the former we convolved the instantaneous evolution with a Gaussian window of 40 fs and normalised all time points to the maximum amplitude of either the finite or instantaneous case for clarity. As expected, the resolution limited case shows deviations from the "true" instantaneous response due to the temporal averaging. Importantly, the molecular evolution over the first 100 fs is qualitatively captured and even pronounced changes on time scales shorter than the temporal resolution are observable. These simulations underline that our trueSMS measurements indeed capture the, temporal resolution limited, ultrafast evolution of the molecule under observation and that changes on time-scales faster than the resolution are a direct result of the dynamical evolution of the system.

**Supplementary Information 4. Fourier transformation of the recorded fluorescence signals**

We extract the fluorescence excitation and true SMS spectra directly as the real part of a fast Fourier transformation. This procedure is justified as our experiment operates in a phase-locked manner (see Supplementary Information 2) as in NMR spectroscopy [29]. In brief, the signal ($S(\Delta t)$) at each effective time delay for a fluorescence excitation experiment is proportional to:

$$S(\Delta t) \propto \int_{-\infty}^{\infty} |\cos[\pi(\nu - \nu_0)\Delta t]|^2 * \sigma_e(\nu)\mathrm{d}\nu,$$

or, for a trueSMS experiment, to:

$$S(\Delta t) \propto S_0 - \int_{-\infty}^{\infty} |\cos[\pi(\nu - \nu_0)\Delta t]|^2 * \sigma_{se}(\nu)\mathrm{d}\nu,$$

with $\sigma_e$ being the fluorescence excitation spectrum, $\sigma_{se}$ the stimulated emission spectrum and $S_0$ the signal in the absence of the probe, or depletion, pulse (Supplementary Figure 6a). These two effective time delay dependent signals behave like cosine functions convolved with the Fourier transformation of the spectral envelope with a phase of either zero or π (Supplementary Figure 6b). The Fourier transformations of these respective functions are both real albeit the former showing a positive and the latter a negative amplitude (Supplementary Figure 6c).

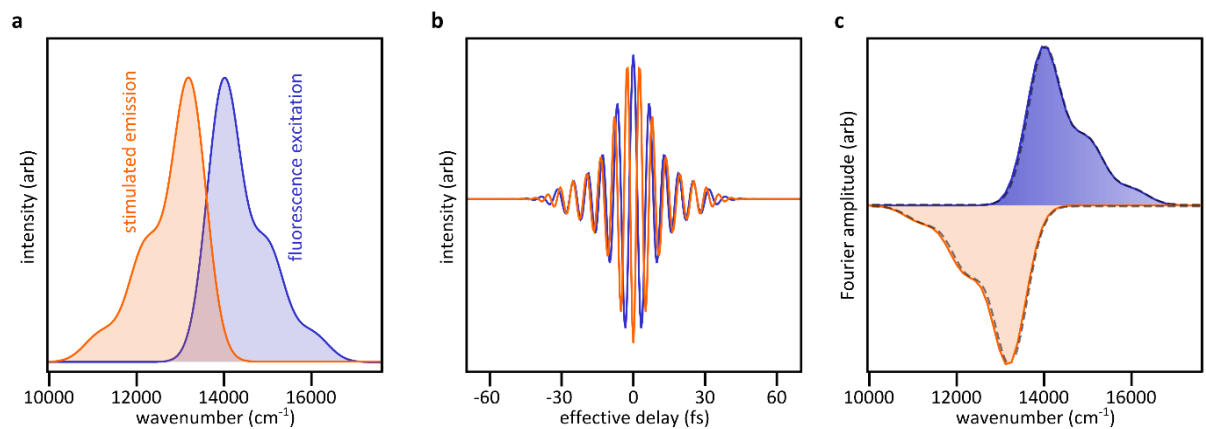

**Supplementary Figure 6: Fourier transformation simulations.** (a) Stimulated emission and fluorescence excitation spectra exhibiting vibronic transitions. (b) Signal obtained by scanning the effective time delay of a square pulse covering the whole spectral range shown in (a) for the fluorescence excitation experiment (blue) and the trueSMS experiment (orange). (c) Real part of the fast Fourier transformations of (b) in comparison to the initially simulated fluorescence excitation spectrum and the inverted stimulated emission spectrum.

**Supplementary Information 5, Pump and probe power dependence**

All experiments presented in the main manuscript are performed in the linear regime of both the excitation and the deexcitation pulses. Supplementary Figure 7 shows the DBT fluorescence scaling as a function of the pump/probe power as well as the position of the powers employed in the experiments presented in Figure 3 and 4 of the main manuscript. These powers correspond to 115 kW/cm$^2$ (pump) and 142 kW/cm$^2$ (probe). The power employed in the fluorescence excitation experiment (Figure 2 in the main manuscript) was a factor of 32 lower which also explains the dramatically reduced fluorescence counts (approximately 2 photos per ms in Figure 2 and approximately 60 photons per ms in Figure 3 of the main manuscript).

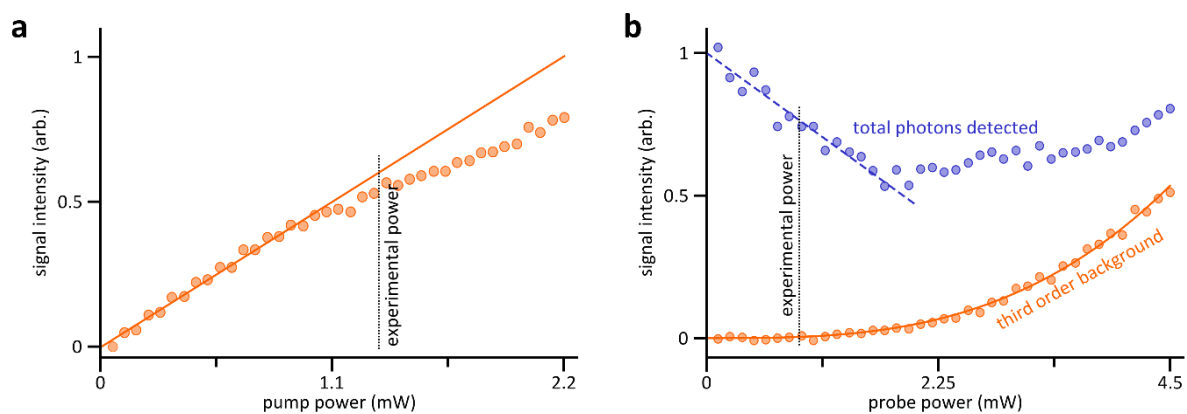

**Supplementary Figure 7: Pump and dump power dependence as well as non-resonant background.** (a) Fluorescence intensity as a function of the pump power (dots) alongside a linear guide to the eye (line). The power employed in the trueSMS experiment is indicated. (b) Fluorescence intensity as a function of the probe power (blue dots) alongside a linear guide to the eye (line). The power employed in the experiment is indicated. The scaling of the nonlinear background obtained from an area containing anthracene (orange dots) is shown in comparison to a cubic fit.

Both pump and probe powers were chosen such that operation in the linear regime was ensured. In the trueSMS experiment it should in principle not be detrimental to operate in the nonlinear regime of the pump pulse as its sole purpose is to populate the excited electronic state. We do not expect to be able to observe nonlinear contributions to the excited state dynamics given our limited sensitivity. Operation at the edge of the linear regime does, however, reduce the possibility of photodamage while simultaneously maximising the amount of fluorescence photons in the detection. We would like to add that we found the almost equally intense probe pulse to be much more phototoxic than the pump pulse even though it is often assumed that deexcitation of a system by means of a dump pulse prevents photobleaching as it removes the molecule from its excited electronic state.

For the modulation pulses (pump pulse in the fluorescence excitation experiment, Figure 2 in the main manuscript) and the probe pulse in the trueSMS experiment (Figure 3 and Figure 4) it is crucial to operate in the linear regime. The spectral modulation leads to a reduction in power which will be translated into a spectral modulation. If the modulation function is nonlinear this transfer will distort the final spectrum. As an extreme case, for a completely saturated transition the spectral modulation would fail to induce any measureable signal modulation hence suggesting that the pulse is not able to excite the molecule.

The nonlinear background observed in Supplementary Figure 7b seems to originate from both the anthracene matrix as well as the coverglass and might be due to spectral broadening or a similar third order process given that our detection window is spectrally extremely close to the probe pulse spectrum. At our excitation level and signal to noise ratio, however, this background does not contribute to our signal (see also Figure 4b in the main manuscript, green line and the respective Fourier transformation).

**Supplementary Information 6, Pulse shaping artefact**

We observe a deviation from zero amplitude in the blue shifted part below 800 nm of the trueSMS spectra of DBT presented in Figure 3d in the main manuscript even though no probe intensity is present in this spectral region. This feature is the result of a slight distortion induced by the finite switching time of the SLM. In brief, every time a new amplitude mask is applied to the SLM the voltages of each pixel in the liquid crystal mask are adjusted to reorient the crystals hence applying the desired modulation. This process takes approximately 100 ms to complete, depending on how big the voltage change has been. In the NIR spectral region we require almost the full voltage range to allow for arbitrary pulse shapes and the switching takes considerably longer. For a full 0 to 1 amplitude switch we observe a change of approximately 95% within 100 ms followed by another 100 ms to complete the process. To keep the duty cycle of our experiment at an acceptable level we decided to start data acquisition 100 ms after switching the SLM based on preliminary experiments with different waiting times that showed that the settling time induced artefacts are within the noise level of a non-averaged experiment as can be seen from the individual spectra presented in both Figure 3 and 4 of the main manuscript. These artefacts are, however, visible if multiple experiments are averaged. We additionally ensured that the switching-induced spectral distortions do not deteriorate the spectral quality within the detection window by comparing scattering spectra recorded from off-resonant with different switching times (100, 200 and 300 ms). Within our noise level all three measurements resulted in equivalent spectra with the latter waiting times not showing the spectral shoulder upon averaging.

1. Musser, A. J. *et al.* Evidence for conical intersection dynamics mediating ultrafast singlet exciton fission. *Nat. Phys.* **11,** 352–357 (2015).

2. Schoenlein, R. W., Peteanu, L. A., Mathies, R. A. & Shank, C. V. The First Step in Vision: Femtosecond Isomerization of Rhodopsin. *Science* **254,** 412–415 (1991).

3. Rose, T. S., Rosker, M. J. & Zewail, A. H. Femtosecond Realtime Observation of Wave Packet Oscillations (Resonance) in Dissociation Reactions. *J. Chem. Phys.* **88,** 6672–6673 (1988).

4. Dostál, J., Pšenčík, J. & Zigmantas, D. Mapping the energy flow through the entire photosynthetic apparatus in situ. *Nat. Chem.* **8,** 705–710 (2015).

5. Shank, C. V. Measurement of ultrafast phenomena in the femtosecond time domain. *Science* **219,** 1027–1031 (1983).

6. Kukura, P., Celebrano, M., Renn, A. & Sandoghdar, V. Single-Molecule Sensitivity in Optical Absorption at Room Temperature. *J. Phys. Chem. Lett.* **1,** 3323–3327 (2010).

7. Chong, S., Min, W. & Xie, X. S. Ground-State Depletion Microscopy: Detection Sensitivity of Single-Molecule Optical Absorption at Room Temperature. *J. Phys. Chem. Lett.* **1,** 3316–3322 (2010).

8. Gaiduk, A., Yorulmaz, M., Ruijgrok, P. V. & Orrit, M. Room-Temperature Detection of a Single Molecule's Absorption by Photothermal Contrast. *Science* **330,** 353–356 (2010).

9. Min, W. *et al.* Imaging chromophores with undetectable fluorescence by stimulated emission microscopy. *Nature* **461,** 1105–1109 (2009).

10. Moerner, W. E. & Kador, L. Optical detection and spectroscopy of single molecules in a solid. *Phys. Rev. Lett.* **62,** 2535–2538 (1989).

11. Maser, A., Gmeiner, B., Utikal, T., Götzinger, S. & Sandoghdar, V. Few-photon coherent nonlinear optics with a single molecule. *Nat. Photonics* **10,** 450–453 (2016).

12. Orrit, M. & Bernhard, J. Single Pentacene Molecules Detected by Fluorescence Excitation in a p-Terphenyl Crystal. *Phys. Rev. Lett.* **65,** 2716 (1990).

13. Nie, S., Chiu, D. T., Zare, R. N. & Zaret, R. N. Probing individual molecules with confocal fluorescence microscopy. *Science* **266,** 1018–1021 (1994).

14. Moerner, W. E. & Orrit, M. Illuminating Single Molecules in Condensed Matter. *Science* **283,** 1670–1676 (1999).

15. van Dijk, E. *et al.* Single-Molecule Pump-Probe Detection Resolves Ultrafast Pathways in Individual and Coupled Quantum Systems. *Phys. Rev. Lett.* **94,** 78302 (2005).

16. Hernando, J. *et al.* Effect of disorder on ultrafast exciton dynamics probed by single molecule spectroscopy. *Phys. Rev. Lett.* **97,** 2–5 (2006).

17. Tian, P. F., Keusters, D., Suzaki, Y. & Warren, W. S. Femtosecond phase-coherent two-dimensional spectroscopy. *Science* **300,** 1553–1555 (2003).

18. Brinks, D. *et al.* Ultrafast dynamics of single molecules. *Chem. Soc. Rev.* **43,** 2476–2491 (2014).

19. Toninelli, C. *et al.* Near-infrared single-photons from aligned molecules in ultrathin crystalline films at room temperature. *Opt. Express* **18,** 6577–6582 (2010).

20. Sanders, J. N. *et al.* Compressed sensing for multidimensional electronic spectroscopy experiments. *J. Phys. Chem. Lett.* **3,** 2697–2702 (2012).

21. Scherer, N. F. *et al.* Fluorescence-detected wave packet interferometry: Time resolved molecular spectroscopy with sequences of femtosecond phase-locked pulses. *J. Chem. Phys.* **95,** 1487 (1991).

22. Albrecht, A. W., Hybl, J. D., Gallagher Faeder, S. M. & Jonas, D. M. Experimental distinction between phase shifts and time delays: Implications for femtosecond spectroscopy and coherent control of chemical reactions. *J. Chem. Phys.* **111,** 10934 (1999).

23. Brinks, D. *et al.* Visualizing and controlling vibrational wave packets of single molecules. *Nature* **465,** 905–8 (2010).

24. Hildner, R., Brinks, D. & van Hulst, N. F. Femtosecond coherence and quantum control of single molecules at room temperature. *Nat. Phys.* **7,** 172–177 (2011).

25. Piatkowski, L., Gellings, E. & van Hulst, N. F. Broadband single-molecule excitation spectroscopy. *Nat. Commun.* **7,** 10411 (2016).

26. Weigel, A., Sebesta, A. & Kukura, P. Shaped and Feedback-Controlled Excitation of Single Molecules in the Weak-Field Limit. *J. Phys. Chem. Lett.* **6,** 4023–4037 (2015).

27. Kennis, J. T. M. *et al.* Ultrafast Protein Dynamics of Bacteriorhodopsin Probed by Photon Echo and Transient Absorption Spectroscopy. *J. Phys. Chem. B* **106,** 6067–6080 (2002).

28. Monmayrant, A., Weber, S. & Chatel, B. A newcomer's guide to ultrashort pulse shaping and characterization. *J. Phys. B-At. Mol. Opt.* **43,** 1–18 (2010).

29. Ernst, R. R., Bodenhausen, G. & Wokaun, A. *Principles of Nuclear Magnetic Resonance in One and Two Dimenisons*. (1990).

30. Mukamel, S. & Biggs, J. D. Communication: Comment on the Effective Temporal and Spectral Resolution of Impulsive Stimulated Raman Signals. *J. Chem. Phys.* **134,** 161101(R) (2011).

31. Polli, D., Brida, D., Mukamel, S., Lanzani, G. & Cerullo, G. Effective Temporal Resolution in Pump-Probe Spectroscopy With Strongly Chirped Pulses. *Phys. Rev. A* **82,** 053809(R) (2010).

32. Nicolet, A. A. L., Hofmann, C., Kol'chenko, M. A., Kozankiewicz, B. & Orrit, M. Single dibenzoterrylene molecules in an anthracene crystal: spectroscopy and photophysics. *ChemPhysChem* **8,** 1215–1220 (2007).

33. Dobryakov, A. L., Kovalenko, S. A. & Ernsting, N. P. Coherent and sequential contributions to femtosecond transient absorption spectra of a rhodamine dye in solution. *J. Chem. Phys.* **123,** 44502 (2005).

34. Cheng, J.-X., Volkmer, A. & Xie, X. S. Theoretical and experimental characterization of coherent anti-Stokes Raman scattering microscopy. *J. Opt. Soc. Am. B* **19,** 1363-1375 (2002).

35. Weigel, A., Sebesta, A. & Kukura, P. Dark Field Microspectroscopy with Single Molecule Fluorescence Sensitivity. *ACS Photonics* **1,** 848–856 (2014).

36. Jonas, D. M. Two-dimensional femtosecond spectroscopy. *Annu. Rev. Phys. Chem.* **54,** 425–463 (2003).

37. Shim, S.-H. & Zanni, M. T. How to turn your pump-probe instrument into a multidimensional

spectrometer: 2D IR and Vis spectroscopies via pulse shaping. *Phys. Chem. Chem. Phys.* **11,** 748–61 (2009).

38. Tekavec, P. F., Lott, G. A. & Marcus, A. H. Fluorescence-detected two-dimensional electronic coherence spectroscopy by acousto-optic phase modulation. *J. Chem. Phys* **127,** 1–21 (2007).

39. Wende, T., Liebel, M., Schnedermann, C., Pethick, R. J. & Kukura, P. Population Controlled Impulsive Vibrational Spectroscopy: Background-and Baseline-Free Raman Spectroscopy of Excited Electronic States. *J. Phys. Chem. A* **118,** 9976–9984 (2014).

40. Liebel, M., Schnedermann, C. & Kukura, P. Vibrationally Coherent Crossing and Coupling of Electronic States During Internal Conversion in β-Carotene. *Phys. Rev. Lett.* **112,** 198302(R) (2014).

41. De, A. K., Monahan, D., Dawlaty, J. M. & Fleming, G. R. Two-dimensional fluorescence-detected coherent spectroscopy with absolute phasing by confocal imaging of a dynamic grating and 27-step phase-cycling. *J. Chem. Phys.* **140,** 194291 (2014).

42. Kane, D. J. & Trebino, R. Charaterization of Arbitrary Femtosecond Pulses Using Frequency-Resolved Optical Gating. *IEEE J. Quantum Elect.* **29,** 571–579 (1993).

43. Lozovoy, V. V., Pastirk, I. & Dantus, M. Multiphoton intrapulse interference. IV. Ultrashort laser pulse spectral phase characterization and compensation. *Opt. Lett.* **29,** 775-777 (2004).

44. von Vacano, B., Buckup, T. & Motzkus, M. Shaper-assisted collinear SPIDER: fast and simple broadband pulse compression in nonlinear microscopy. *J. Opt. Soc. Am. B* **24,** 1091-1100 (2007).

45. Liebel, M., Schnedermann, C., Wende, T. & Kukura, P. Principles and Applications of Broadband Impulsive Vibrational Spectroscopy. *J. Phys. Chem. A* **119,** 9506–9517 (2015).